\documentclass[twocolumn]{aastex62}
\usepackage{amsmath}

\received{}
\revised{}
\accepted{}

\shorttitle{The First Day in the Life of a Magnetar}
\shortauthors{{\c S}a{\c s}maz Mu{\c s} et al.}

\begin{document}

\title{The First Day in the Life of a Magnetar: Evolution of the Inclination Angle, Magnetic Dipole Moment and Braking Index of Millisecond Magnetars During Gamma-Ray Burst Afterglows}

\correspondingauthor{Sinem {\c S}a{\c s}maz Mu{\c s}}
\email{sasmazmus@itu.edu.tr}

\author[0000-0002-9669-5825]{Sinem {\c S}a{\c s}maz Mu{\c s}}
\affil{\.{I}stanbul Technical University \\
Faculty  of Science  and  Letters, \\
Physics Engineering  Department, \\
34469,  \.{I}stanbul, Turkey}

\author[0000-0003-1002-659X]{Sercan {\c C}{\i}k{\i}nto{\u g}lu}
\affil{\.{I}stanbul Technical University \\
Faculty  of Science  and  Letters, \\
Physics Engineering  Department, \\
34469,  \.{I}stanbul, Turkey}

\author[0000-0003-2044-5026]{U{\u g}ur Ayg{\"u}n}
\affil{\.{I}stanbul Technical University \\
Faculty  of Science  and  Letters, \\
Physics Engineering  Department, \\
34469,  \.{I}stanbul, Turkey}

\author[0000-0003-3670-6059]{I.~Ceyhun Anda{\c c}}
\affil{\.{I}stanbul Technical University \\
Faculty  of Science  and  Letters, \\
Physics Engineering  Department, \\
34469,  \.{I}stanbul, Turkey}

\author[0000-0001-5999-0553]{K.~Yavuz Ek\c{s}i}
\affil{\.{I}stanbul Technical University \\
Faculty  of Science  and  Letters, \\
Physics Engineering  Department, \\
34469,  \.{I}stanbul, Turkey}




\begin{abstract}
The afterglow emission of some gamma-ray bursts (GRBs) show a shallow decay (plateau) phase implying continuous injection of energy. The source of this energy is very commonly attributed to the spin-down power of a nascent millisecond magnetar. The magnetic dipole radiation torque is considered to be the mechanism causing the spin-down of the neutron star. This torque has a component working for the alignment of the angle between rotation and magnetic axis, i.e.,\ inclination angle, which has been neglected in modelling GRB afterglow light curves. Here, we demonstrate the evolution of the inclination angle and magnetic dipole moment of nascent magnetars associated with GRBs. We constrain the initial inclination angle, magnetic dipole moment and rotation period of seven magnetars by modelling the seven long-GRB afterglow light curves. We find that, in its first day, the inclination angle of a magnetar decreases rapidly. The rapid alignment of the magnetic and rotation axis may address the lack of persistent radio emission from mature magnetars. We also find that in three cases the magnetic dipole moments of magnetars decrease exponentially to a value a few times smaller than the initial value. The braking index of nascent magnetars, as a result of the alignment and magnetic dipole moment decline, is variable during the afterglow phase and always greater than three.
\end{abstract}


\section{Introduction} \label{sec:intro}
The idea that a millisecond magnetar, i.e.\ a neutron star with a super-strong magnetic field ($B \sim 10^{13} - 10^{15}\,{\rm G}$) \citep{dun92}, may be formed during a GRB has long been suggested \citep{uso92,dun92}. With the launch of the Swift telescope, different segments in many GRB light curves have been identified \citep{zha+06,nou+06}. The ``plateau" segment is suggested to be powered by the spin-down of a nascent millisecond magnetar and frequently modelled by employing the spin-down power of a magnetar \citep[see, e.g.,][]{dai98,fan06,las+17}.

The radiation emitted by a rotating neutron star has been considered to be due to its magnetic dipole moment and studied in the framework of the magnetic dipole braking model \citep{gol68,pac68}. For a neutron star rotating in vacuum and neglecting the alignment component, this model predicts a spin-down relation of the form $\dot{\Omega} = - K \Omega^3$ where $\Omega$ is the angular velocity of the star, $K \equiv 2(\mu\sin\alpha)^2 /(3Ic^3)$, $c$ is the speed of light, $\mu$ is the magnetic dipole moment of the star, $I$ is the moment of inertia and $\alpha$ is the inclination angle. From this one can solve the spin-down luminosity, $L_{\rm sd}=-I \Omega \dot{\Omega}$, as $L_{\rm sd} = L_0 (1 + t/t_0)^{-2}$ where $L_0 = I K \Omega_{0}^{4}$ and $t_0 = \left(2 K\Omega_0^2 \right)^{-1}$. This leads to $L_{\rm sd} \propto t^{-2}$ at late time. If the second derivative of the angular velocity ($\ddot{\Omega}$) is measured, the braking index $n \equiv \ddot{\Omega}\Omega / \dot{\Omega}^{2}$ can be determined. The value of the braking index is used for the assessment of the pulsar spin-down mechanisms and models. Magnetic dipole braking model predicts a constant braking index of $n=3$. 

Recently, \citet{las+17} \citep[see also][]{lu+19} have invoked the somewhat more general spin-down relation $\dot{\Omega} \propto - \Omega^n$ for modelling the spin-down of putative magnetars in two \textit{short} GRBs. This model involves the assumption that the braking index is constant and has the solution $L_{\rm sd} = L_0 (1 + t/t_0)^{-(n+1)/(n-1)}$. From the fits to the afterglow light curves of two \textit{short} GRBs, GRB~130603B and GRB~140903A, \citet{las+17} found the value of the braking indices of the putative magnetars as $n = 2.9\pm 0.1 $ and $n=2.6 \pm 0.1$, consistent with the measured braking indices of conventional pulsars. For ordinary pulsars with ages of $10^3 - 10^4$~yr the assumption of constant braking index may be reasonable for an episode of 10 days. The light curves of GRB afterglows, on the other hand, from the first second to 10 days, amount to more than 5 decades of data and assuming a constant braking index is not warranted for these cases.

Moreover, the alignment of the inclination angle due to the magnetic dipole torque acting on a neutron star is expected as the system will evolve in a way to reduce the spin-down losses \citep{phi+14}, but is usually neglected for ordinary pulsars on the grounds that its progress would be slowed-down, for non-spherical pulsars, by dissipation due to the presence of a solid crust \citep{gol70a}. It is not possible to justify the neglect of the alignment component for nascent magnetars taking role in GRBs as their crust may not have solidified yet as also suggested by \citet{lan18}. \citet{xu15} studied the effect of rapid linear change in inclination angle on the light curves of GRB afterglows. 

In this work we model the X-ray afterglow light curves of selected GRBs (\S \ref{sec:grbsamp}) with the spin-down power of a nascent neutron star: $L_{\rm X} = \eta L_{\rm sd}$ where $\eta$ is an efficiency factor for conversion of spin-down power to X-ray luminosity. We assume that the star is slowing down in the presence of a corotating plasma, under the action of the alignment torque coupled with the spin-down torque (\S \ref{sec:model}). When necessary we allowed also for the exponential evolution of the magnetic dipole moment, either growth or decay. We inferred the period, inclination angle, magnetic dipole moment of the star at the start of the plateau phase, evolutionary time scale of the magnetic dipole moment and its relaxed value by following a Bayesian framework (\S \ref{sec:parest}) and calculated the braking indices of nascent magnetars.

\begin{figure*}[ht!]
\figurenum{1.1}
\includegraphics[scale=0.9]{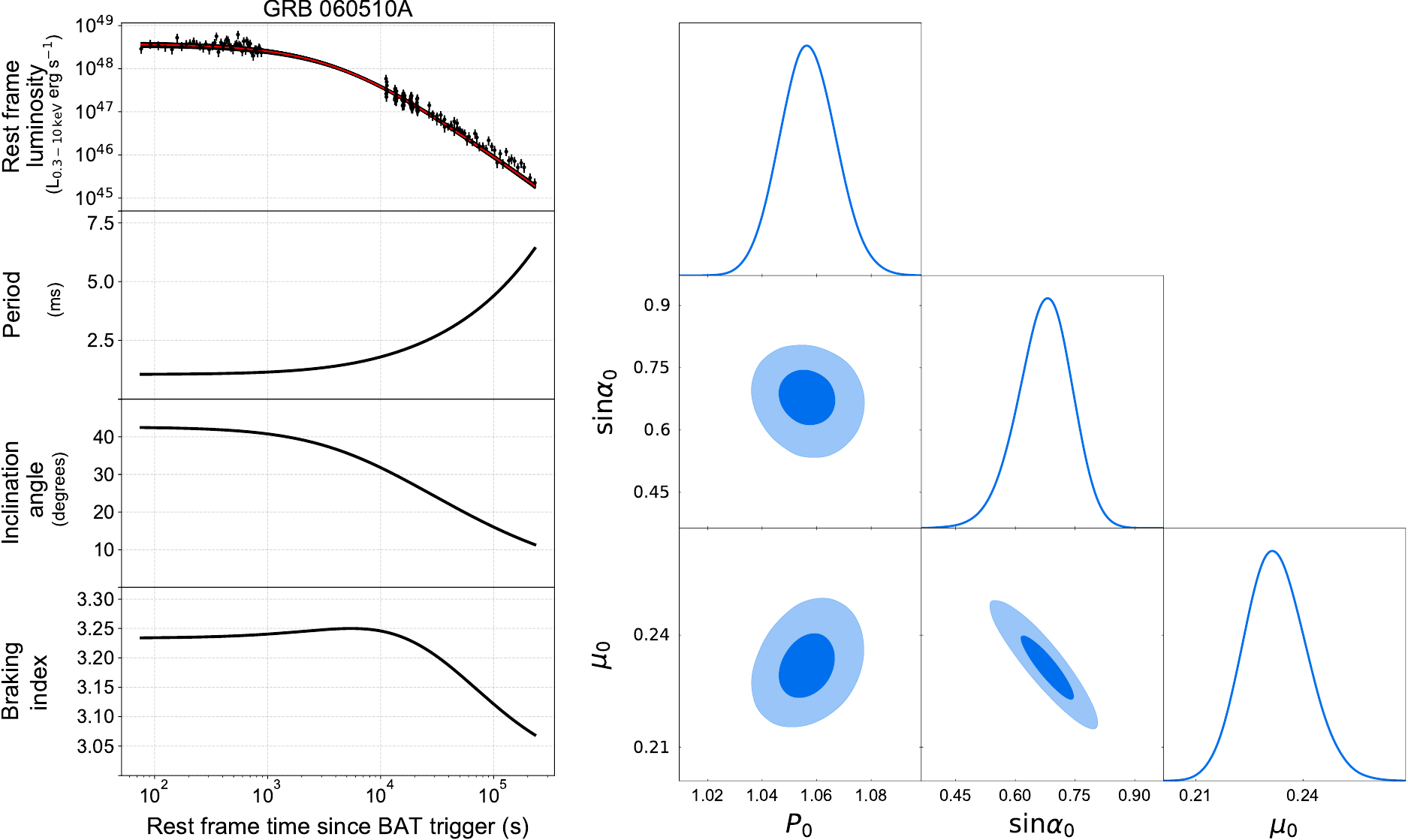}
\caption{\textbf{Evolution of putative nascent magnetar parameters and triangle plot for GRB 060510A.} \textbf{Left panel:} Luminosity, period, inclination angle and braking index evolution of the putative nascent magnetar in GRB 060510A for the constant magnetic dipole moment case. Solid black lines in the luminosity evolution panel represent 500 models randomly selected from the posterior probability distribution and the red line represents the median value of all samples. \textbf{Right panel:} 1D and 2D posterior probability distributions of the parameters. Contours indicate 1 and 2 sigma levels. \label{fig:060510A}}
\end{figure*}

\section{Methods} \label{sec:method}

\subsection{Model equations} \label{sec:model}
The spin-down \citep{spi06} and alignment \citep{phi+14} torques in the presence of a corotating plasma \citep{gol69} are 
\begin{align}
I \frac{{\rm d} \Omega}{{\rm d}t} &= 
	-\frac{ \mu^2 \Omega^3}{c^3} (1 + \sin^2 \alpha ) 
\label{eq:mdr1}  \\
I \frac{{\rm d} \alpha}{{\rm d}t} &= 
	-\frac{ \mu^2 \Omega^2}{c^3} \sin \alpha \cos \alpha 
\label{eq:mdr2}.
\end{align}
The alignment in the presence of the corotating plasma is slower compared to the vacuum case \citep{phi+14}. In the plasma-filled magnetosphere model the braking index is given as \citep{arz+15} 
\begin{equation}
n = 3 + 2 \left( \frac{\sin \alpha \cos \alpha}{1 + \sin^2 \alpha}\right)^2
\label{eq:n0}
\end{equation} 
The magnetic dipole moment of nascent magnetars could be changing in the time-frame of our analysis either because the magnetic field \citep[see, e.g.,][]{dun92} or the radius \citep{bur86} of a millisecond nascent magnetar could change. The evolution of the magnetic field within the star is governed by diffusive type partial differential equation and depends on complicated microscopic physics of the crust  \citep[see, e.g.,][]{urp+92}. To simplify matters we assume that this initial transient stage can be described by the ordinary differential equation
\begin{equation}
\dot{\mu} = - K (\mu - \mu_{\infty})^{q + 1},
\end{equation} 
where $K$ and $q$ are constants and $\mu_{\infty}$ is the value the magnetic moment tends to relax.  The long term decay of the magnetic field \citep{dun92,col+00} with a time scale of $10^{2}-10^{4}$ yr is beyond the scope of this paper and can not be constrained with the short-term ($\lesssim 10$ days) data we use in this work. It means that $\mu_{\infty}$ which we assume constant in the time-frame we are interested actually changes slowly.
This nonlinear differential equation has the solution
\begin{equation}
\mu = \mu_{\infty} + (\mu_0 - \mu_{\infty})\left( 1 + q \frac{t}{t_{\rm m}}\right)^{-1/q}
\label{eq:nonlinear}
\end{equation}
where $\mu_0$ is the initial magnetic moment, and 
\begin{equation}
t_{\rm m}=(\mu_0 - \mu_{\infty})^{-q}/K
\end{equation}
is the field decay time-scale. Note that at the linearity limit 
($q \rightarrow 0$) the solution becomes
\begin{equation}
\mu = \mu_{\infty} + 
		(\mu_0 - \mu_{\infty}) \exp(-t/t_{\rm m}).
\label{eq:linear}
\end{equation}

In case the magnetic dipole moment evolves simultaneously with the inclination angle, the braking index is given as \citep{eks17} 
\begin{equation}
n = 3 + 
	2 \left( \frac{\sin \alpha \cos \alpha}{1 + \sin^2 \alpha}\right)^2 + 
	4 \frac{\tau_{\rm c}}{\tau_{\mu}}
\label{eq:n1}
\end{equation}
where $\tau_{\rm c} \equiv - \Omega/2\dot{\Omega}$ 
and $\tau_{\mu} \equiv - \mu/\dot{\mu}$. Using Equation~(\ref{eq:linear}) we obtain 
\begin{equation}
\tau_{\mu} = \frac{t_{\rm m}}{\mu_{\infty}/\mu - 1}
\end{equation}
and by using Equation (\ref{eq:mdr1})
\begin{equation}
\tau_{\rm c} = \frac{I c^3}{2\mu^2 \Omega^2 (1+\sin^2 \alpha)}.
\end{equation}
According to the model employed in this work, the braking index of a magnetar tends to relax to $n=3$, the canonical theoretical value for spinning-down magnetic dipoles, as the second and the third terms in \autoref{eq:n1} vanish in time. By this we do not imply that mature magnetars will have $n=3$ as there could be other processes, such as winds \citep{gao+16}, ambipolar diffusion and Hall drift \citep{gol+92} leading to deviations from the simple magnetic dipole spin-down law in those cases. The processes that may dominate the spin-down following the afterglow stage is beyond the scope of this work.
We assume that the isotropic equivalent X-ray luminosity of the afterglow scales with the 
spin-down luminosity $L_{\rm sd} = - I \Omega \dot{\Omega}$ so that
\begin{equation}
L_{\rm X} = \eta\frac{ \mu^2 \Omega^4}{c^3} (1 + \sin^2 \alpha), 
\label{eq:Lx}
\end{equation}
where $\eta$ is a constant. This factor not only represents how efficiently the spin-down luminosity is converted into X-rays in the observed band, but also takes care of the beaming hence can be greater than unity.

\subsection{Parameter estimation} \label{sec:parest}
Our parameter sets are \{$P_{0}$, $\sin\alpha_{0}$, $\mu_{0}$\} for the constant magnetic dipole moment case and \{$P_{0}$, $\sin \alpha_{0}$, $\mu_{0}$, $\mu_{\infty}$, $t_{\rm m}$\} for the changing magnetic dipole moment case. In order to estimate the nascent magnetar parameters and their credible regions we followed the Bayesian approach \citep[see, e.g.,][]{siv06} with the posterior probability distribution defined as 
\begin{equation}
\textmd{posterior} \propto \textmd{likelihood} \times \textmd{prior}.
\end{equation}

The Swift/XRT light curve production phase has a set of criteria in order to achieve Gaussian statistics \citep{eva+07}. Therefore, it is reasonable to assume error distribution is Gaussian. Here, we used a Gaussian ln-likelihood function\footnote{In practice it is suitable to work with natural logarithm of the probability distributions (see, e.g., \citet{hog18})} such that 
\begin{equation}
\textmd{ln}\left({\textmd{likelihood}}\right) = 
-\frac{1}{2}\sum_{i=1}^{N}\left[\frac{\left(y_{i} 
- f(x_{i},\theta)\right)^2}{\sigma_{y,i}^2} 
+ \textmd{ln}\left(2\pi\sigma_{y,i}^2\right) \right]
\label{lnlike}
\end{equation}
where $N$ is the number of data points, $y_{i}$ and $\sigma_{y,i}$ represents ith data point and its uncertainty, respectively. $f(\theta)$ is the model with a set of $\theta$ parameters corresponding to $x_{i}$ abscissa value. We used uniform prior probability in specified parameter ranges (see Table \ref{tab:priors}) defined as
\begin{equation}
\textmd{ln}\left(\textmd{prior}\right)=\begin{cases}0, & \text{if $\theta_{l}<\theta<\theta_{u}$}.\\
-\infty, & \text{otherwise}.
\end{cases}
\label{eq:prior}
\end{equation}

We do not have any strong prior information for the value of the parameters. Therefore, we choose to use uniform distribution. The upper and lower limits of the parameters ($\theta_l$ and $\theta_u$ values in Equation~(\ref{eq:prior})  are chosen such that they cover large enough parameter region for a nascent millisecond magnetar. The lower value of period is chosen above the breakup limit \citep[see, e.g.,][]{coo+94}. Lower values of magnetic dipole moments correspond to typical magnetic field values for normal pulsars (10$^{12}$ G). Upper values correspond to field values an order of magnitude greater than the highest inferred dipole field values of mature magnetars. The magnetic dipole moment decay timescale range is chosen to cover variability from $\sim$10 seconds to 100 days. We think this range is large enough for a short term rapid evolution of magnetic field decay that we consider in our model. We increased the upper values of parameters  (except the inclination angle) an order of magnitude (two orders of magnitude for the decay timescale) and found that our results are not affected.

The luminosity model, $f(\theta)$, we employed is given in Equation~(\ref{eq:Lx}). We solved the coupled equations, (\ref{eq:mdr1}) and (\ref{eq:mdr2}), described in Section~\ref{sec:model} using \texttt{scipy.integrate.odeint} to find $\Omega(t)$ and $\alpha(t)$ and calculated the light curve with Equation~(\ref{eq:Lx}). 
The efficiency factor $\eta$ in Equation (\ref{eq:Lx}) is not a parameter that could be determined independent of $\mu_0$ and $P_0$. Because of the underlying symmetry of the equations, $P_0 \rightarrow P_0\sqrt{\eta}$ and $\mu_0 \rightarrow \mu_0\sqrt{\eta}$ will result with the same light curve. 
The X-ray efficiency factors obtained from observations for mature pulsars range between $10^{-5} - 10^{-1}$ \citep[see, e.g.,][]{kar+12}. On the other hand, the jet correction factor if we consider the relativistic beaming of the radiation can reach up to $\sim$ 500 \citep[see, e.g.,][]{fra+01}.
Therefore, we do our analysis for $\eta=1$. However, using the underlying symmetry above one can easily estimate the $P_0$ and $\mu_0$ values that would give the same results for a different value of $\eta$. In Figure~\ref{fig:eta}, we present the possible values of $P_0$ and $\mu_0$ for different values of $\eta$ and mark the values that correspond to $\eta=0.1$ and $\eta=10$ as well as $\eta=1$.
Although we fixed the moment of inertia to its canonical value, i.e., $I=10^{45}$~g~cm$^2$, in all calculations, $P_0$ and $\mu_0$ for different $I$ values will transform under $P_0 \rightarrow P_0\sqrt{I_{45}}$ and $\mu_0  \rightarrow~\mu_0 I_{45}$ where $I_{45}\equiv I/(10^{45}$~g~cm$^2)$. We present possible values of $P_0$ and $\mu_0$ as $I_{45}$ changes for $\eta=1$ in Figure \ref{fig:eta} and specifically mark $I_{45} = 0.5$ and $I_{45} = 3.0$ values since these values represent approximately lower and upper values given by calculations based on different equations of state \citep[see, e.g.,][]{hae+07}.

\begin{center}
\begin{deluxetable}{lccc}[h!]
\tablecaption{Parameter Boundaries for Uniform Prior Distributions. \label{tab:priors}}
\tablecolumns{4}
\tablehead{
\colhead{Parameters} &
\colhead{Lower Limit} &
\colhead{Upper Limit} 
}
\startdata
$P_{0}$ (ms)                          & $0.70$    & $10^{2}$ \\
$\sin \alpha_{0}$                     & $10^{-3}$ & $0.99$ \\
$\mu_{0}$ ($10^{33}$ G cm$^{3}$)      & $10^{-3}$ & $10$ \\
$\mu_{\infty}$ ($10^{33}$ G cm$^{3}$) & $10^{-3}$ & $10$ \\
$t_{\rm m}$ (day)                     & $10^{-4}$ & $10^{2}$ \\
\enddata
\end{deluxetable}
\end{center}

In order to calculate the posterior probability distributions of our parameter sets, i.e., \{$P_{0}$, $\sin \alpha_{0}$, $\mu_{0}$\} and \{$P_{0}$, $\sin \alpha_{0}$, $\mu_{0}$, $\mu_{\infty}$, $t_{\rm m}$\}, we used an affine-invariant Markov Chain Monte-Carlo Ensemble sampler, \texttt{emcee} \citep{emcee13,emceerc218}. We used 500 walkers for each parameter. In the first run of the \texttt{emcee} we used $50,000$ steps for burn-in phase. In the main phase we adjusted the number of steps according to the integrated autocorrelation time ($\tau_{f}$) \citep{goo10,emcee13} which is calculated using \texttt{acor} \citep{acor12} package. At every hundred steps we calculated $\tau_{f}$ and its difference from the previous step. When the step number is smaller than $100\tau_{f}$ and the differences in new and old $\tau_{f}$ value is smaller than $0.01$, we assumed that the chain has converged\footnote{\url{https://emcee.readthedocs.io/en/latest/tutorials/monitor/}}. In order to be more conservative we continued sampling until either the step number is reached up to five hundred times the maximum $\tau_{f}$ or $100,000$ steps. Then we ``thinned'' the sample by the half of the mean $\tau_{f}$. In the second run of the \texttt{emcee} we initialized the chains around a small Gaussian ball of most probable values of the parameters determined in the first run of the \texttt{emcee}. The parameter values and their uncertainties are obtained from the median and standard deviation of the posterior probability distributions. Finally, we calculated the evolution of the braking index from Equation (\ref{eq:n0}) and Equation (\ref{eq:n1}) for constant and changing magnetic dipole moment cases, respectively. The 1D and 2D posterior probability distributions of the parameters are plotted with the \texttt{getdist} \citep{getdist18} package (i.e., triangle plot).

\begin{figure*}[ht!]
\figurenum{2.1}
\includegraphics[scale=0.9]{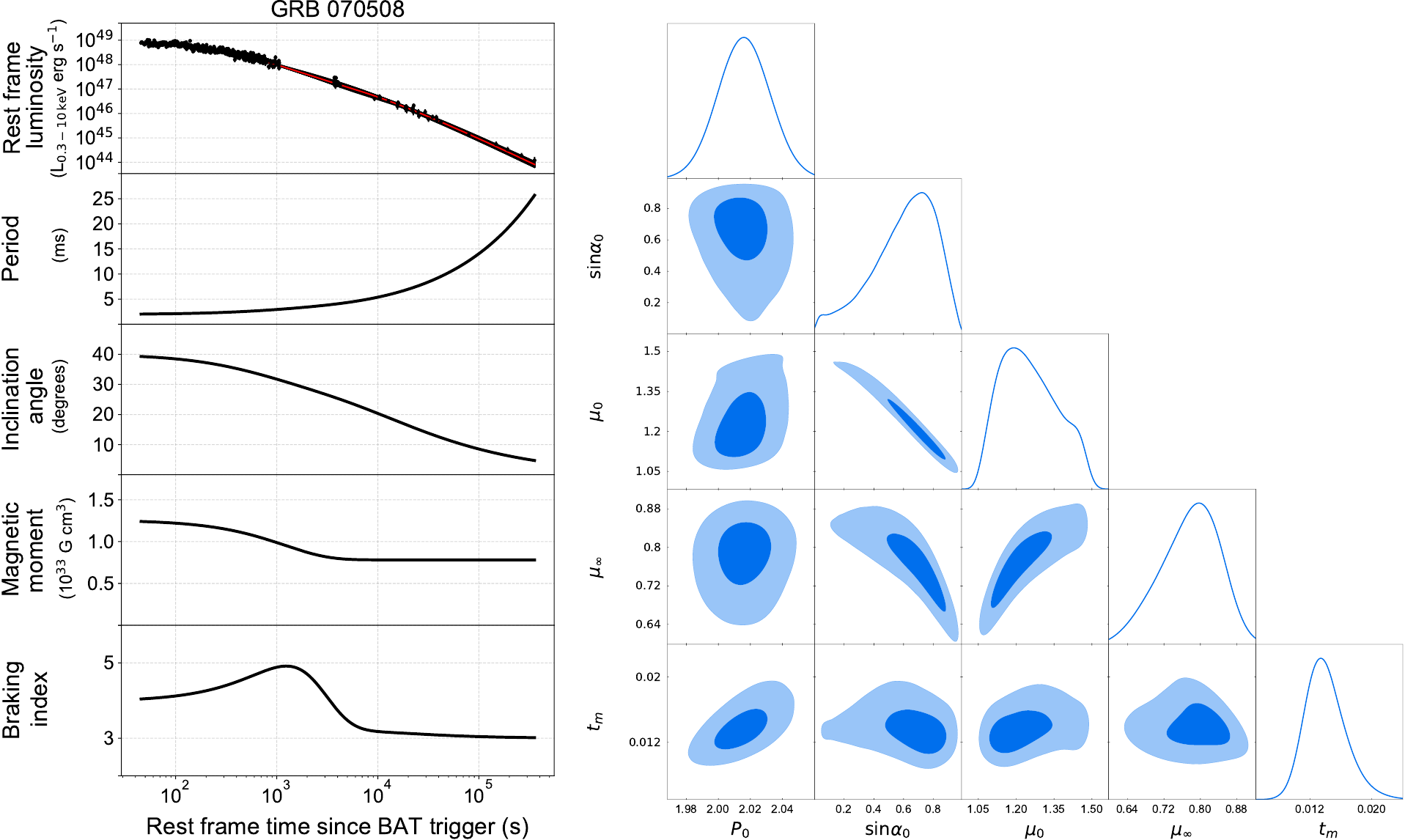}
\caption{\textbf{Evolution of putative nascent magnetar parameters and triangle plot for GRB 070508.} \textbf{Left panel:} Luminosity, period, inclination angle, magnetic dipole moment and braking index evolution of the putative nascent magnetar in GRB 070508 for the changing magnetic dipole moment case. Solid black lines in the luminosity evolution panel represent 500 models randomly selected from the posterior probability distribution and the red line represents the median value of all samples. \textbf{Right panel:} 1D and 2D posterior probability distributions of the parameters. Contours indicate 1 and 2 sigma levels. \label{fig:070508}}
\end{figure*}

\subsection{GRB sample} \label{sec:grbsamp}
GRBs display a bimodal distribution of duration \citep{kou+93}: the short- ($t < 2$ s) and long-duration bursts ($t \gtrsim 2$ s). Short GRBs have long been considered, on theoretical grounds, to arise from the merger of two neutron stars \citep{bli84,eic+89}, an idea which  was spectacularly confirmed by the detection of gravitational waves \citep{abb+17a} released during inspiral of two neutron stars associated  with GRB~170817 \citep{abb+17b}. The merger of binary neutron stars may result with the formation of a rapidly rotating ($P \sim 1\,{\rm ms}$) magnetar \citep{dun92} which is centrifugally supported and may collapse to a black hole upon slow-down. The origin of the long-duration GRBs are understood as the deaths of massive stars in core-collapse supernovae of Type 1c \citep{woo93}. Some long duration GRBs, nevertheless, are believed to be associated with magnetars (see, e.g.,\ \citet{gom17}). Therefore, we do not use the prompt emission duration as a selection criteria for candidate searches since both types could host magnetars.
\begin{center}
\begin{deluxetable}{lcc}[h!]
\tablecaption{GRB Sample and Its Parameters \label{tab:grbsamp}}
\tablecolumns{3}
\tablewidth{0pt}
\tablehead{
\colhead{GRB} &
\colhead{\hspace{1.0cm}Redshift\tablenotemark{\rm a}\hspace{1.0cm}} &
\colhead{\hspace{0.7cm}Photon Index} \\
\colhead{} & \colhead{\hspace{1.0cm}($z$)\hspace{1.0cm}} &
\colhead{\hspace{0.7cm}$\Gamma$}
}
\startdata
060510A & \hspace{1.0cm}1.2$^{\rm b}$  & \hspace{0.7cm}$1.81\pm 0.08$ \\
070420  & \hspace{1.0cm}0.66$^{\rm c}$ & \hspace{0.7cm}$1.778\pm 0.095$ \\
070521  & \hspace{1.0cm}2.0865         & \hspace{0.7cm}$1.89\pm 0.13$ \\
140629A & \hspace{1.0cm}2.275          & \hspace{0.7cm}$1.96\pm 0.11$ \\
070508  & \hspace{1.0cm}0.82           & \hspace{0.7cm}$1.79\pm 0.135$ \\
091018  & \hspace{1.0cm}0.971          & \hspace{0.7cm}$2.0\pm 0.115$ \\ 
161219B & \hspace{1.0cm}0.1475         & \hspace{0.7cm}$1.83\pm 0.06$ \\
\enddata
\tablenotetext{\rm a}{Redshift and photon index values 
are taken from the Swift \textit{XRT} GRB light curve 
repository \citep{eva+07,eva+09} unless otherwise stated.}
\tablenotetext{\rm b}{\cite{oat+12}}
\tablenotetext{\rm c}{\cite{xia11}}
\end{deluxetable}
\end{center}

\begin{deluxetable*}{lcccc}
\tablecaption{Estimated Values of the Putative Nascent Magnetar Parameters for Constant Magnetic Dipole Moment Case. The values of $P_0$ and $\mu_0$ depend on the choice of value of $\eta$ and $I_{45}$. The results given here are obtained for $\eta=I_{45}=1$. In Section \ref{sec:parest} we show how to convert these values for other choices of $\eta$ and $I_{45}$, and present them in Figure \ref{fig:eta}. \label{tab:cnstB}}
\tablecolumns{5}
\tablewidth{0pt}
\tablehead{
\colhead{GRB} &
\colhead{$P_{0}$} &
\colhead{$\sin \alpha_{0}$} &
\colhead{$\mu_{0}$} &
\colhead{$\chi^{2}/{\rm dof}$} \\
\colhead{} &
\colhead{(ms)} &
\colhead{} &
\colhead{($10^{33}$~G\,cm$^{3}$)} &
\colhead{} 
}
\startdata
060510A & $1.057\pm 0.010$ & $0.675 \pm 0.069$ &  $0.232 \pm 0.009$ & $163.17/159$ \\
070420  & $3.916 \pm 0.040$ & $0.623 \pm 0.092$ & $1.012 \pm 0.049$ & $178.87/142$ \\
070521  & $1.357 \pm 0.024$ & $0.508 \pm 0.201$ & $0.699 \pm 0.055$ & $81.62/82$ \\
140629A & $1.951 \pm 0.027$ & $0.586 \pm 0.172$ & $1.149 \pm 0.085$ & $66.96/89$ \\
\enddata
\end{deluxetable*}

We obtained the $0.3-10$ keV unabsorved flux values of the GRBs from the Swift \textit{XRT} GRB light curve repository\footnote{\url{http://www.swift.ac.uk/xrt_curves/}} \citep{eva+07,eva+09} including flux values obtained from photon counting (PC) mode and also from windowed timing (WT) mode if available. We computed the luminosity values from flux light curves using
\begin{equation}
L = 4 \pi d_{\rm L}^{2}(z) F_{\rm X} k(z)
\end{equation}
Here, $d_{\rm L}(z)$ is the luminosity distance which depends on the redshift of the source. $F_{\rm X}$ is the unabsorbed flux values. $k$ is the cosmological $k$-correction due to cosmological expansion \citep{blo+01} which is given as
\begin{equation}
k(z) = (1 + z)^{(\Gamma - 2)}
\end{equation}
Here $\Gamma$ is the photon index of the power law. We obtained the photon index values from Swift \textit{XRT} GRB lightcurve repository. We calculated the luminosity distance using a flat $\Lambda$CDM cosmological model with cosmological parameters $H_{0} = 71 \ {\rm km}\ {\rm s}^{-1} \ {\rm Mpc}^{-1}$ and $\Omega_{M} = 0.27$ with \texttt{astropy.cosmology} subpackage \citep{astropy18}. We obtained most of the redshift values of the sources from the same GRB light curve repository. For the sources which there is no redshift information in that repository, we used published values from literature. We list these values and their references for the candidates presented in this work in Table \ref{tab:grbsamp}. 

Exploring all the available GRBs we focused on the ones which contain a ``plateau'' phase. The presence of this ``plateau'' phase is what rekindled the millisecond magnetar model \citep{zha+06,nou+06}, so we assumed those events would be the most probable candidates for hosting a magnetar as the central engine. Yet, our model has limits, e.g.\ does not involve the physics of fallback accretion that is expected to form in core collapse \citep[see e.g.,][]{col71} and debris disks expected to form in the aftermath of the merger of two neutron stars \citep{ros07}, so we do not expect it to fit all the light curves. We thus eliminated many afterglows with complicated light curves that can not be addressed with the physics involved in here. In order to decide whether the light curve can be fitted with the model, we employed an initial non-linear least squares fit using lmfit package \citep{lmfit14}. Thus, the subset of the GRBs modelled here likely do not have dynamically significant discs that could apply a torque comparable with the magnetic dipole torques. Here we present the strongest candidates, GRBs 060510A, 070420, 070521, 140629A, 070508, 091018 and 161219B, for which the magnetar central engine would be favoured. We note that although we did not use the duration of GRBs as a selection criterion, we find that all strong candidates belong to long-GRB class \citep{060510a+t90,070420+t90a,070420+t90b,070508+t90,070521+t90a,070521+t90b,140629a+t90,161219b+t90a,161219b+t90b,091018+t90}.

\begin{center}
\begin{deluxetable*}{lcccccc}
\tablecaption{Estimated Values of the Putative Nascent Magnetar Parameters 
for Changing Magnetic Dipole Moment Case. The values of $P_0$ and $\mu_0$ depend on the choice of value of $\eta$ and $I_{45}$. The results given here are obtained for $\eta=I_{45}=1$. In Section \ref{sec:parest} we show how to convert these values for other choices of $\eta$ and $I_{45}$, and present them in Figure \ref{fig:eta}. \label{tab:expB}}
\tablecolumns{7}
\tablewidth{0pt}
\tablehead{
\colhead{GRB} & 
\colhead{$P_{0}$} & 
\colhead{$\sin\alpha_{0}$} &
\colhead{$\mu_{0}$} &
\colhead{$\mu_{\infty}$} &
\colhead{$t_{\rm m}$} &
\colhead{$\chi^{2}/{\rm dof}$} \\
\colhead{} & 
\colhead{(ms)} &
\colhead{} &
\colhead{($10^{33}$~G\,cm$^{3}$)} &
\colhead{($10^{33}$~G\,cm$^{3}$)} &
\colhead{(days)} &
\colhead{}
}
\startdata
070508 & $2.016 \pm 0.016$ & $0.633 \pm 0.213$ & $1.241 \pm 0.110$ & $0.782 \pm 0.065$ & $0.014 \pm 0.003$ & $426.59/483$ \\
091018  & $2.922 \pm 0.039$ & $0.523 \pm 0.236$ & $1.804 \pm 0.170$ & $0.683 \pm 0.063$ & $0.016 \pm 0.002$ & $139.23/134$ \\
161219B & $7.088 \pm 0.055$ & $0.617 \pm 0.085$ & $0.507 \pm 0.024$ & $0.178 \pm 0.010$ & $0.281 \pm 0.018$ & $660.70/407$ \\
\enddata
\end{deluxetable*}
\end{center}

\section{Results and Discussion}
\label{sec:results}
\begin{figure*}[ht!]
\centering
\figurenum{1.2}
\includegraphics[scale=0.8]{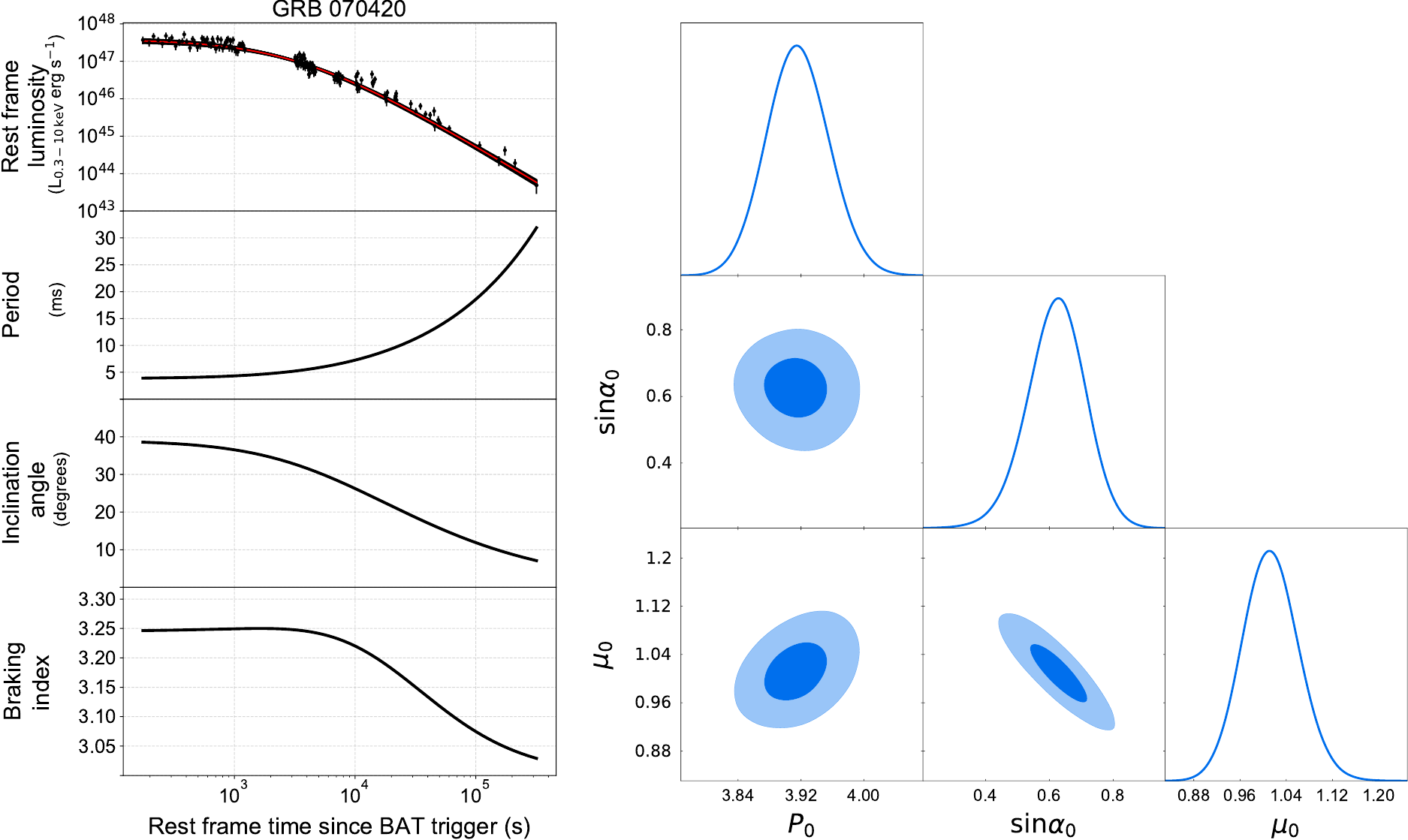}
\caption{Same as \autoref{fig:060510A} but for GRB 070420. \label{fig:070420}}
\end{figure*}

\begin{figure*}
\centering
\figurenum{1.3}
\includegraphics[scale=0.8]{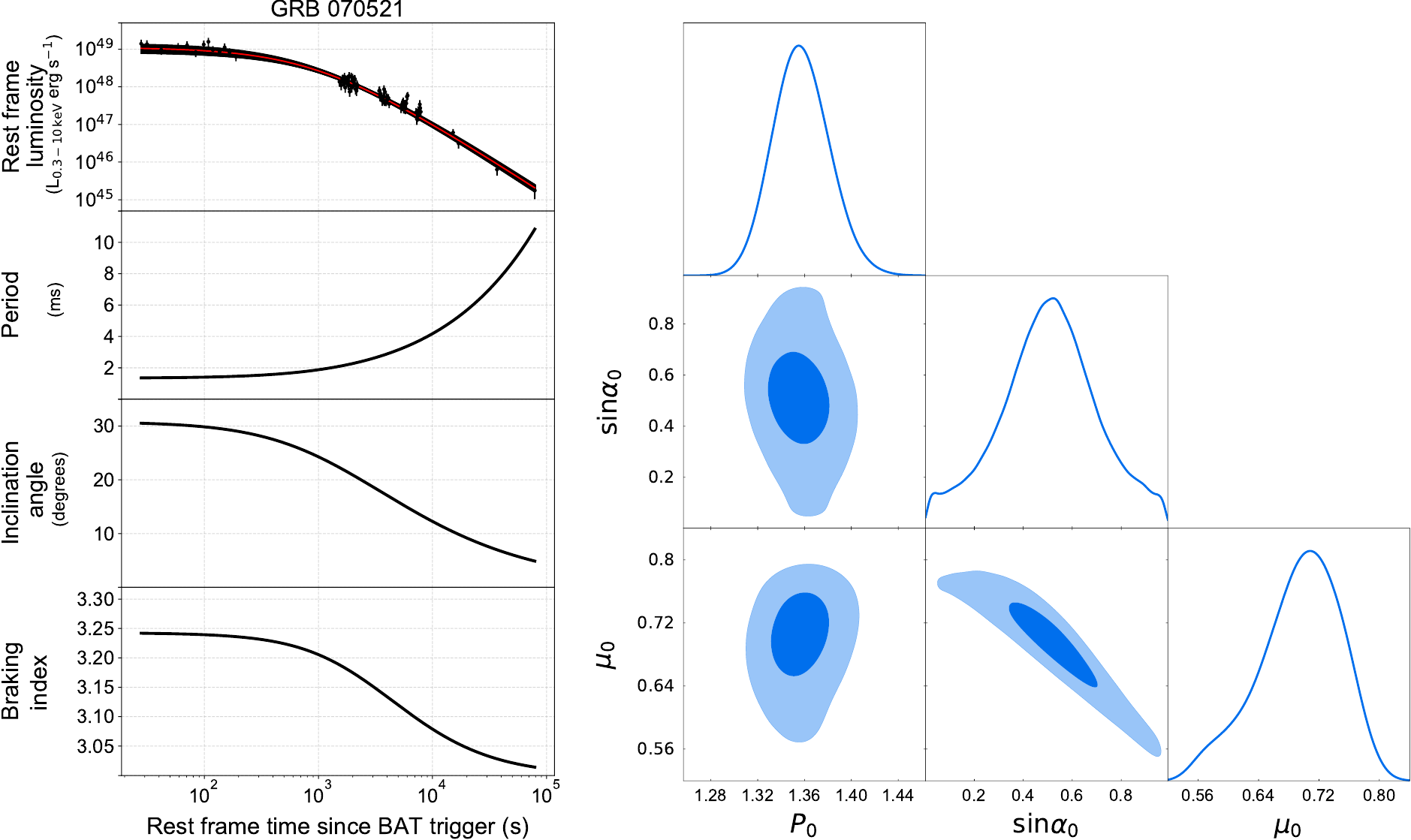}
\caption{Same as \autoref{fig:060510A} but for GRB 070521. \label{fig:070521}}
\end{figure*}

\begin{figure*}
\centering
\figurenum{1.4}
\includegraphics[scale=0.9]{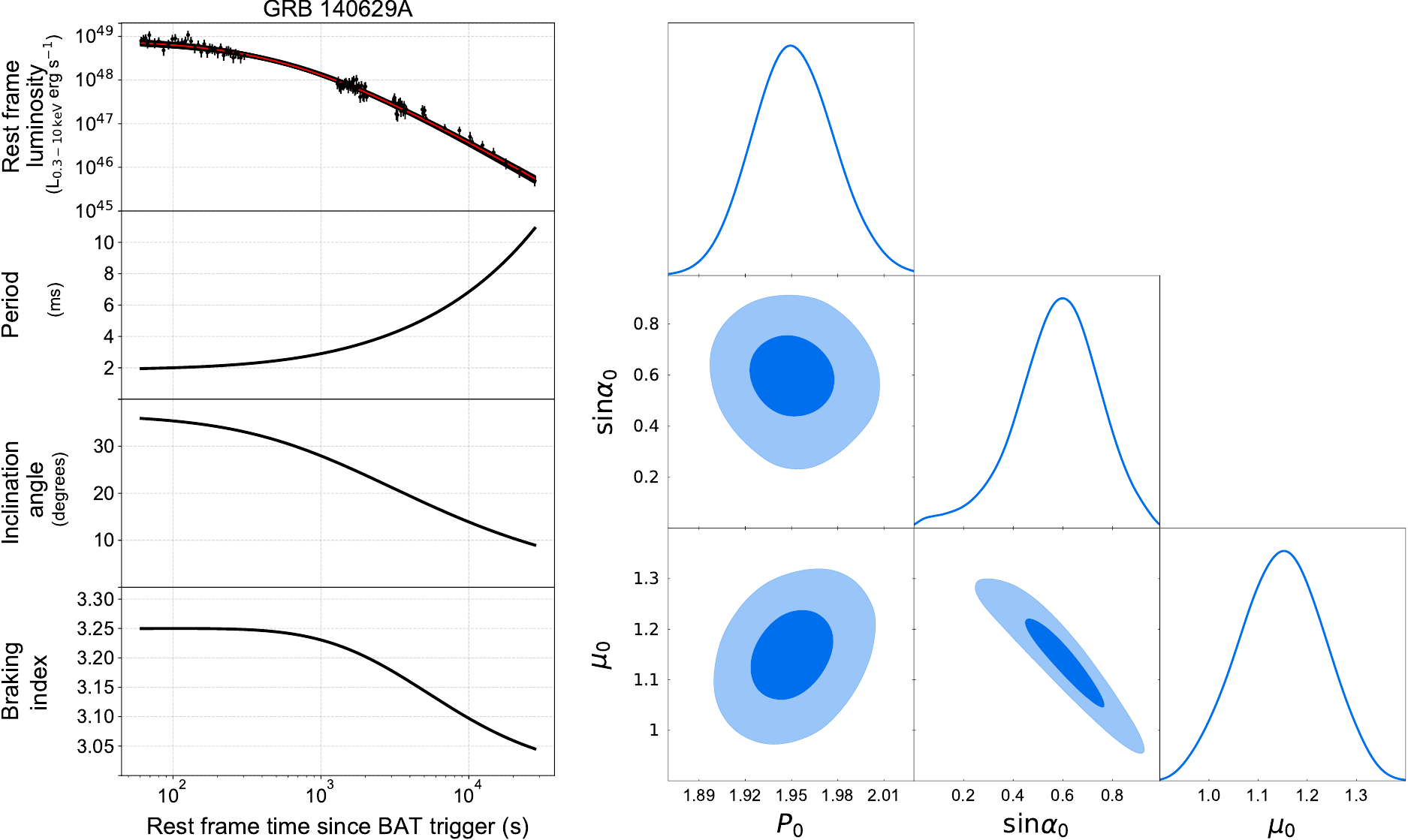}
\caption{Same as \autoref{fig:060510A} but for GRB 140629A. \label{fig:140629A}}
\end{figure*}

We modelled the X-ray afterglow light curves of seven long GRBs with magnetic dipole radiation model described in Section \ref{sec:model}. Four GRBs in our sample, GRBs 060510A, 070420, 070521 and 140629A, are best modelled with a constant magnetic dipole moment (see Table~\ref{tab:cnstB}). 
We present the evolution of luminosity, period, inclination angle and braking index of these GRBs in Figures \ref{fig:060510A}, \ref{fig:070420}, \ref{fig:070521} and \ref{fig:140629A} (left panels).

\begin{figure*}
\centering
\figurenum{2.2}
\includegraphics[scale=0.9]{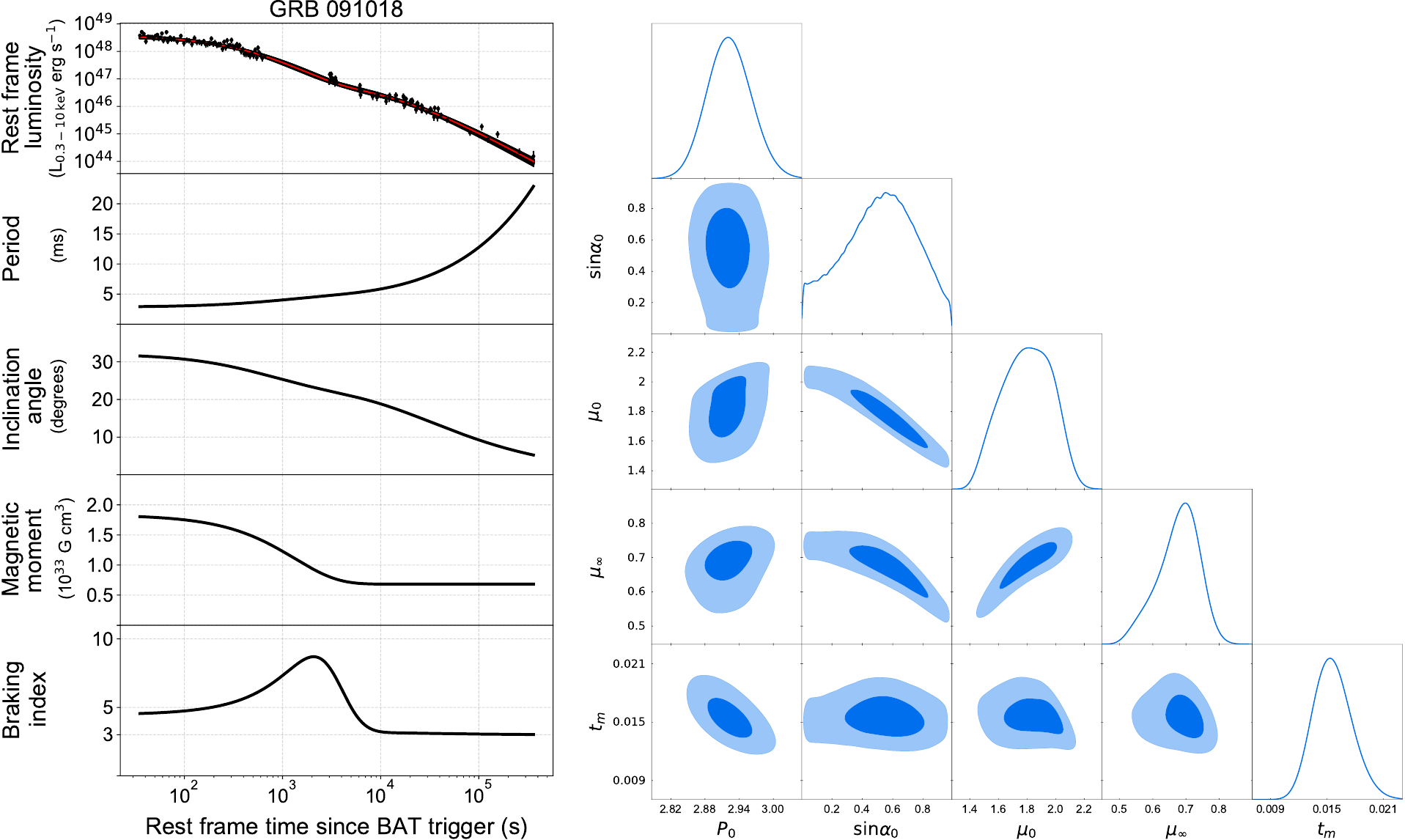}
\caption{Same as \autoref{fig:070508} but for GRB 091018. \label{fig:091018}}
\end{figure*}

\begin{figure*}
\centering
\figurenum{2.3}
\includegraphics[scale=0.9]{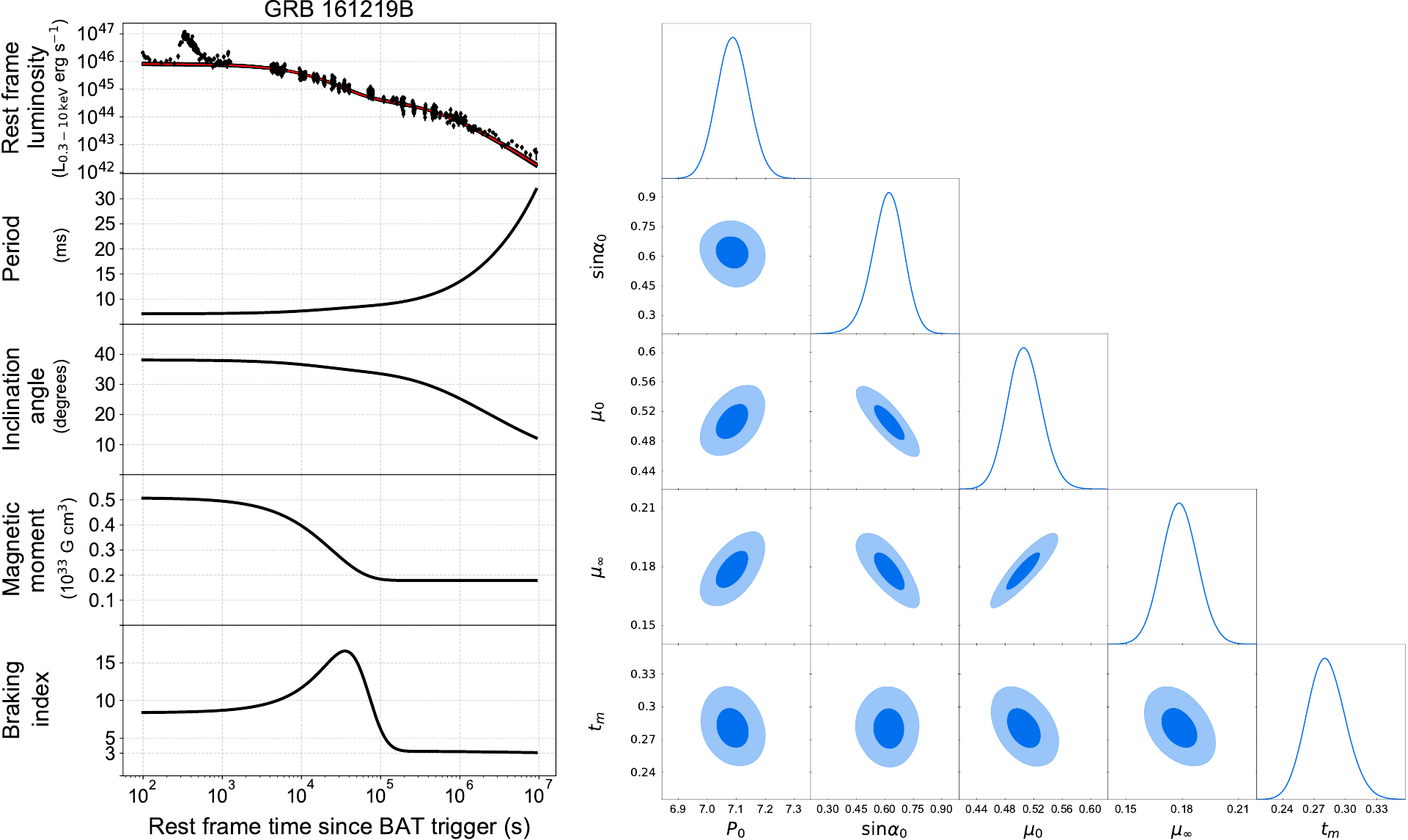}
\caption{Same as \autoref{fig:070508} but for GRB 161219B. \label{fig:161219B}}
\end{figure*}

In three cases, GRBs 070508, 091018 and 161219B, the model with constant magnetic dipole moment resulted in reduced chi-squared values greater than $\sim 3$. Consequently, for these GRBs we allowed the evolution of the magnetic dipole moment according to Equation \eqref{eq:linear}. We find that for these GRBs magnetic dipole moments of the magnetars decrease rapidly. The putative nascent magnetar parameters including the evolutionary time-scale of the magnetic dipole moment and its relaxed value for the three GRBs are listed in Table~\ref{tab:expB} and evolution of the parameters is presented in Figures \ref{fig:070508}, \ref{fig:091018} and \ref{fig:161219B} (left panels). 
1D and 2D posterior probability distributions of the parameters for all GRBs are shown on the right panel of each figure.

We also carried out an analysis with the power-law with a lower value model presented in Equation (\ref{eq:nonlinear}) keeping $q$ free for our GRB samples with changing magnetic dipole moments, i.e., GRBs 070508, 091018 and 161219B. Although this model has some support with respect to the exponential model it has two drawbacks: (i) our fits give substantially different $q$ values for each source; (ii) the inclination angle can not be constrained well if $q$ is a free parameter. We observed that with fixed $q$ values in the power-law with bottom value model, the inclination angle can be constrained better as $q$ goes to zero. Thus, in the lack of a unique $q$ value supported by a theory we present our results only for the exponential model $q=0$ for its simplicity. We observed that different values of $q$ lead to small changes in the values of the fit parameters but it does not change our general conclusion that the inclination angle and the magnetic field changes at this early stage of magnetars.    

The drop in the magnetic dipole moment, $\mu=\frac12 BR^3$, might either be due to a decaying magnetic field, $B$, or the radius of the star, $R$, settling. However, the time-scale for the settling of the radius of a proto-neutron star is $\sim10\,{\rm s}$ \citep{bur86} which is much shorter than the time-scale for the evolution of the magnetic moment inferred in this work, $0.02-0.3~{\rm day}$. It is then unlikely that the change in the magnetic dipole moment is due to the change in the radius of the star.  The time-scale for the change in the magnetic dipole moment inferred from the fits is much shorter than the secular magnetic field decay time-scale in the magnetar model \citep[see e.g.][]{col+00} $\sim10^4$ yr. What we ``observe'' here could be the tail of the evolution of the initial strong $B \sim 10^{16}\,{\rm G}$ magnetic field associated with the prompt emission \citep{ben+17}.

It is interesting to see that for the 7 GRBs we analysed the inclination angles of nascent magnetars at the start of the plateau phase are found to have values distributed in the narrow range of $\sim 30^\circ-45^\circ$. As a result of the alignment between the rotation and the magnetic axis, the spin-down luminosity, $L$ declines less rapidly than $L \propto t^{-2}$ resulting from pure spin-down. In the model employed in this work the putative magnetars in GRBs are required to have slightly higher magnetic dipole moments (compared to the previous models that do not allow for alignment) to address a certain light curve. 
\begin{figure*}
\figurenum{3}
\begin{center}
\includegraphics[scale=0.95]{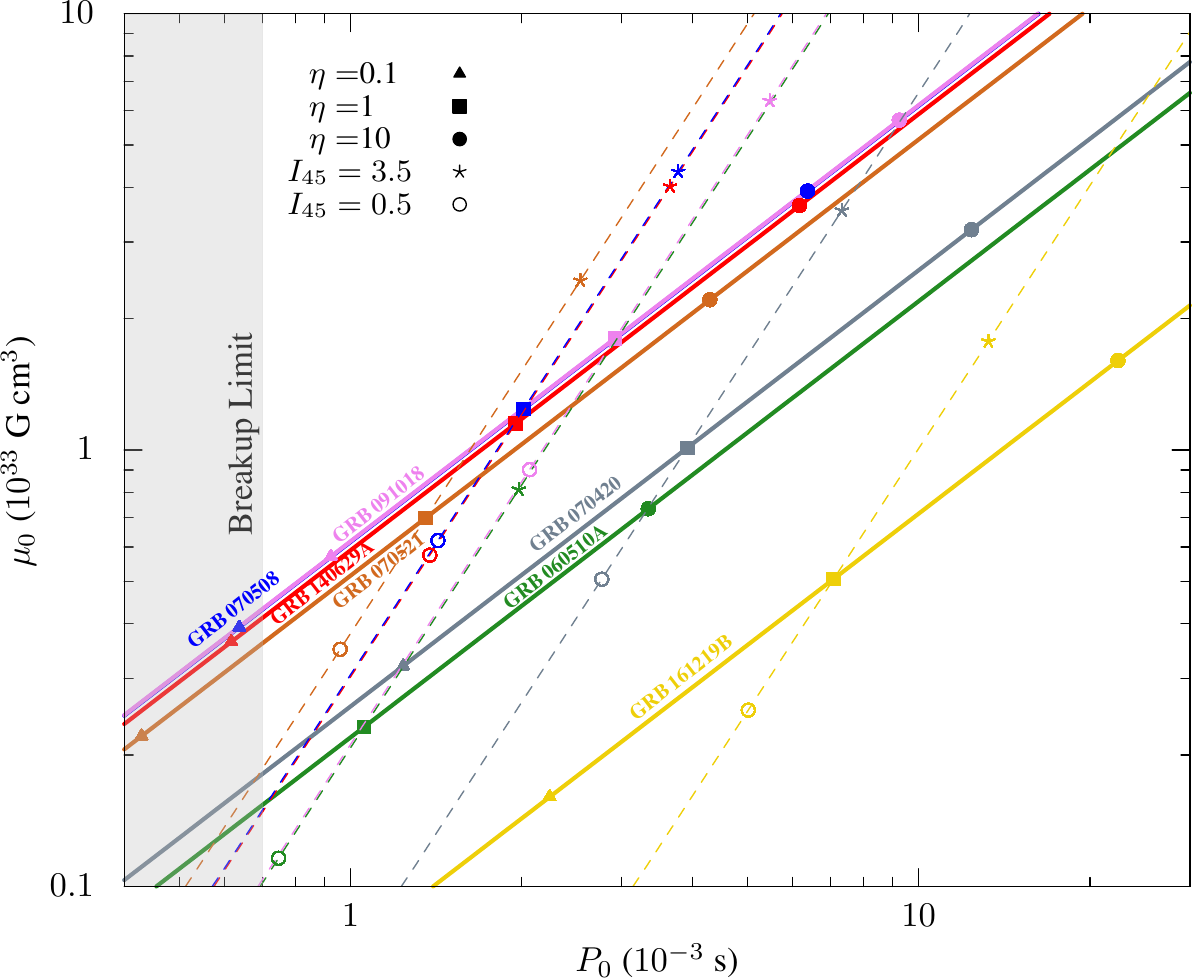}
\caption{Possible values of $P_{0}$ and $\mu_{0}$ for different values of $\eta$  (solid lines) and $I_{45}$ (dashed lines). Each object is represented by two curves (colored the same in the online version) which cross at the $\eta=I_{45}=1$ point shown with a filled square. The gray region shows the range of breakup periods ($\sim$0.4$-$0.7 ms) calculated by \cite{coo+94} for different equations of state.}
\end{center}
\label{fig:eta}
\end{figure*}

Note that all the $P_0$, $\mu_0$ curves (see Figure \ref{fig:eta}) remain below the canonical $P_0=1\,{\rm ms}$, $\mu_0 = 1\times 10^{33}\,{\rm G\,cm^3}$ point for the choice of $I_{45}=1$. It is remarkable that the light curves could be modelled by assuming slower rotating ($P_0\sim10\,{\rm ms}$) magnetars with very strong magnetic dipole moments ($\mu_0\sim10^{34}\,{\rm G\,cm^3}$) depending on the value of $\eta$ and $I_{45}$.

In the model employed here, the inclination angle approaches to small values as $\alpha \propto (t/t_0)^{-1/2}$ in a spin-down time-scale \citep{phi+14}. \citet{mal14} inferred by using polarization angle variation measurements of magnetars that are detected in the radio band and showed that the inclination angles of mature magnetars are small. This could be the consequence of the rapid alignment stage we discuss in this work.

The alignment of the inclination angle between the \textit{dipole} field and the rotation axis may explain the lack of \textit{persistent} radio emission from mature magnetars \citep[e.g.][]{bur+06}. The \textit{transient} radio emission from these magnetars \citep{cam+06} likely result from the quake-triggered twisted current flowing magnetosphere \citep{got+19,wan+19} or \textit{multipole} fields \citep[see e.g.][]{kra+07}. In this picture the distinction between classical magnetars like 1E~2259$+$586 and high magnetic field radio pulsars like PSR~J1119--6127 is that the dipole field of the latter is, somehow, not aligned with the rotation axis.

The same alignment torque would also lead to the formation of very young neutron stars with aligned conventional ($B \sim 10^{12}$~G) magnetic fields though in a longer time scale. Such objects, however, can not show up as rotationally powered pulsars. This may explain the lack of detection of a rotationally powered pulsar in SN~87A \citep{alp+18} and the central compact object in Cas A \citep{cha+01}. This picture, in order to address the existence of many young rotationally powered pulsars, such as Crab, requires that the inclination angle that is reduced during the first years following birth should increase in the longer term. There is indeed evidence that inclination angle of the Crab pulsar has been increasing at a rate $0.62^\circ \pm 0.03^\circ$ per century \citep{lyn+13}. The cause of this counter-alignment is yet to be clarified.

\acknowledgments
We thank anonymous referee for constructive comments. This work made use of data supplied by the UK Swift Science Data Centre at the University of Leicester (http://www.swift.ac.uk/xrt{\_}curves/). S\c{S}M acknowledges post-doctoral research support from \.{I}stanbul Technical University. S\c{S}M and KYE acknowledges support from T\"{U}B\.{I}TAK, The Scientific and Technological Research Council of Turkey, with the project number 118F028.

\vspace{5mm}
\facility{Swift}


\software{\texttt{Astropy} \citep{astropy18},  
          \texttt{Matplotlib} \citep{hun07},
          \texttt{SciPy} \citep{scipy01},
          \texttt{emcee} \citep{emcee13, emceerc218}, 
          \texttt{getdist} \citep{getdist18},
          \texttt{acor} \citep{acor12},
          \texttt{gnuplot} \citep{gnuplot13}
          }

\clearpage

\end{document}